
%
%
\input phyzzx.tex

\tolerance400
\font\bigbf=cmbx10 scaled\magstep2
\def\RG{{\it RG\/} }
\def\SM{{\it SM\/} }

\def \k{\kappa}
\def\ik{\k^{-1}}
\def\il{\k^{-2}}
\def\as{\alpha_3}
\def \l{\lambda}

\def\npb{{Nucl.\ Phys.\ }{\bf B}}
\def\physrep{Phys.\ Reports\ }
\def\plb{Phys.\ Lett.\ }
\def\prd{{Phys.\ Rev.\ }{\bf D}}

\def \p{\phi}
\def \m{\mu}
\def\pd{\partial}
\def\b{\beta}

\line{\hfil LTH 288}
\line{\hfil UM-TH-92-21}
\line{\hfil}
\line{\hfil}
\vfil
{\bf

\line{\hfil THE EFFECTIVE POTENTIAL AND THE RENORMALISATION GROUP\hfil}
}
\bigskip
\bigskip
\line{\hfil\bf C. Ford, D.R.T. Jones, P.W. Stephenson\hfil}
\line{\hfil\it DAMTP, University of Liverpool, Liverpool L69 3BX, UK\hfil}
\medskip
\line{\hfil\it and\hfil}
\medskip
\line{\hfil\bf M.B. Einhorn \hfil}
\line{\hfil\it Randall Laboratory, University of Michigan, Ann Arbor, MI
48109-1120, USA\hfil}
\vskip5cm

\vfil
\line{\bf \hfil Abstract\hfil}
We discuss renormalisation group improvement of the effective
potential both in general and in the context of $O(N)$ scalar
$\p^4$ and the Standard Model. In the latter case we find that
absolute stability of the electroweak vacuum implies that
$m_H\geq 1.95m_t-189~GeV$, for $\as (M_Z) = 0.11$. We point out that
the lower bound on $m_H$ {\it decreases\/} if $\as (M_Z)$ is increased.

\vfill\eject

\line{\bigbf 1. Introduction.\hfil}

\line{\hfill}
The effective potential $V(\p)$ plays a crucial role in determining the
nature of the vacuum in weakly coupled field theories, as was
emphasised in the classic paper of Coleman and Weinberg({\it CW.})
\Ref\cw{E.Weinberg and S.Coleman, \prd7 (1973) 1888} The
loopwise perturbation expansion of $V$ is reliable only for a limited
range of $\p$ ; but, as was recognised by \it CW\rm, it is possible to
extend the range of $\p$ by exploiting the fact that $V$ satisfies a
renormalisation group (\RG) equation. It is therefore possible
to show that in massless $\l\p^4$, $V(\p)$ has a local minimum at
$\p=0$, while  massless scalar  QED has a local minimum for
$\p\not=0$.

Let us review briefly how $V$ is calculated in perturbation
theory, using the (functionally derived) elegant method of Jackiw.
\Ref\Jackiw{R.Jackiw, \prd9 (1974) 1686}
In general one shifts scalar fields: $\p(x) \rightarrow \p + \p_q(x)$,
where $\p$ is $x$-independent. Then $V(\p)$ is given by the sum of vacuum
graphs with $\p$-dependent propagators and vertices. It is not immediately
obvious from this algorithm what the result for the one loop calculation
is; partly for this reason, some authors have preferred to consider graphs
with one $\p_q$-leg, which, it is easy to show, lead to a determination of
${\partial V/ \partial \p}$. All this is very familiar; not so well known,
perhaps, is the following point. Jackiw's algorithm in conjunction with a
specific subtraction scheme ( such as $MS$ or $\overline{MS}$ ) leads to
an expression
for $V(\p)$ such that $V(0)$ is well defined and calculable: and also,
of course generally ignored. Our point is that unless $V(0)$ is
{\sl specifically\/} subtracted (or otherwise dealt with,) then $V(\p)$
fails to satisfy a \RG equation of the usual form. This fact
was noted, for example, in~
\REF\EJ{M.B.Einhorn and D.R.T.Jones, \npb211 (1983) 29} Ref.~\EJ ,
but has often been overlooked, leading to incorrect
``solutions" to the \RG equation. This happens because the form of the
solution transmogrifies the apparently trivial $V(0)$ term into
a $\p$-dependent quantity. We will see how this comes about in sections (2)
and (3) where we discuss various strategies for dealing with $V(0)$, and
their consequences. We will also argue that it is in fact simpler to use the
\RG equation for ${\partial V/ \partial \p}$, since this leads to
an ``improved" form of $V$ that removes the necessity of considering
``improvement" of $V(0)$.

In subsequent sections we explore various forms for the \RG equation
for both $V$ and ${\partial V /\partial \p}$ for various field theories.
We consider in detail scalar $\p^4$ theory, with particular emphasis
on the impact of infra-red divergences on the domain of validity of the
solution. We also consider the standard model, where the behaviour of
$V$ at large $\p$ is important since it can affect the stability of the
electroweak vacuum. Here we improve (in principle) on previous treatments
\REFS\Shers{M.J.Duncan, R. Phillipe and M.Sher, \plb153B (1985) 165;
 M.Sher and H.W.Zaglauer, \plb206B (1988) 527; M.Lindner, M.Sher and
H.W.Zaglauer, \plb228B (1989) 139;
J.Ellis, A.Linde and M.Sher, \plb252B (1990) 203}
\REFSCON\Sher{M.Sher, \physrep179 (1989) 274}\refsend
by our use of a correct form of the \RG solution, and also by use
of a correct form of the 2-loop ${\b}$-function for the Higgs self coupling
\Ref\fij{C.Ford, I.Jack and D.R.T.Jones, \npb (in press)}
; but, as is easy to anticipate, the analysis of Ref.~\Sher~should not be
materially affected. Interestingly, however, we find dependence on
$\as (M_Z)$ that differs significantly from that given in \REF\Sherz{
M.Lindner et al., ref.(4)}~Ref.~\Sherz.

\line{\hfill}
\line{\bigbf 2. The renormalisation group equation  for V.\hfil}

\line{\hfill}
In what follows we consider the \RG equation in renormalisable field
theories with a single renormalisation scale $\mu$ and couplings $\l_i$ of
dimension $\delta_i$. Thus the set $\l_i$ consists of all masses and coupling
constants, both dimensionless and dimensionful. In general, $V$ is a function
$V(\mu, \lambda_i,\p^a)$ where $\p^a$ represents all the scalar fields.
In many cases, however, symmetries may be exploited so that $V$ may be
calculated as a function of a single field $\p$. This is the case in the
standard model, for example. In more complicated cases (involving
supersymmetry, for instance) one frequently chooses to explore a specific
direction in
$\p$-space. Of course ultimately one must then be able to argue that the
absolute minimum of $V$ is indeed in the chosen direction.(This is
not always a trivial matter.\Ref\fjr{J.M.Fr\`ere, D.R.T.Jones and S.Raby,
\npb222 (1983) 11})~In any event, we
will assume for simplicity that it is sufficient to consider the case of
a single $\p$-field only.

It is straightforward to derive the \RG equation satisfied by $V$, but there
is one subtlety. If we calculate $V$ according to the procedure outlined in the
previous section, then the result $\hat V(\mu,\lambda_i,\phi)$
 is such that $\hat V(\mu,\lambda_i,0)$ is a non trivial
function that receives contributions from all orders in perturbation theory.
(In fact $\hat V(\mu,\lambda_i,0)$
 may well have an imaginary part if $\phi=0$ is not a local
minimum of the tree potential, but let us imagine for the moment that this
problem does not arise). Thus we may write
$$
\hat V(\mu,\lambda_i,\phi)=\hat V(\mu,\lambda_i,0)-\sum_{n=1}^{\infty}
{1\over{n!}}\phi^n\Gamma^{(n)}(p_i=0)\eqno(2.1)
$$
where $\Gamma^{(n)}$ represents the 1PI
 Green's function with n $\phi$-legs and all
external momenta set equal to zero. Then by virtue of the \RG equation
satisfied by $\Gamma^{(n)}$, we have
$${\cal D}\hat V-\gamma\phi{\partial \hat V\over{\pd \phi}}={\cal D}\Omega
\eqno(2.2)$$
where we have denoted $\hat V(\mu,\lambda_i,\phi)$ by
$\hat V  $ and $\hat V(\mu,\l_i,0)  $ by
$\Omega$. The operator $\cal D   $ is
$${\cal D}=\m{\pd\over{\pd \m}}+\b_i{\pd\over{\pd \l_i}}.\eqno(2.3)
$$
$\Omega$ is simply a contribution to the vacuum energy on which,
outside of gravity, no observable can depend. Accordingly, we can
make a $\phi$-independent shift in $V$, ie.
$\hat V\rightarrow V=\hat V+\Omega^\prime(\m,\l_i) $
then by choosing $\Omega^\prime $
so that
$${\cal D}\Omega^{\prime}+{\cal D}\Omega=0\eqno(2.4)$$
we can arrange that
$${\cal D}V-\gamma\p{\pd V\over{\pd\p}}=0\eqno(2.5)$$
which is the usual \RG equation for the effective potential.
Thus the \RG equation restricts the form of the ``cosmological
constant" $\Omega+\Omega^\prime$ and leads to observable consequences
when we presently consider \RG ``improvement" of $V$.

On the assumption that we want a potential that satisfies eq.(2.5), then
what is the appropriate choice of $\Omega^\prime?$ The obvious choice is of
course
$$\hbox{i)}\qquad \Omega^\prime=-\Omega.\eqno(2.6)$$
This was advocated, for example, in Ref.~\EJ . Its defect, however, is that as
mentioned above $V$ may have an imaginary part at the origin. A suitable
generalisation to the case when the minimum of $V$ lies at non-zero $\phi$
is given by\Ref\ejt{M.B.Einhorn and D.R.T.Jones, preprint NSF-ITP-92-105}
$$\hbox{ii)}\qquad \Omega^{\prime}=-\hat V(\p)\Bigm\vert_{\p=v}\eqno(2.7)$$
where $v$ is the value of $\phi$ at its minimum. (If $V$ has more than one
local
minimum then any one will give a well defined $V$ satisfying eq.(2.5)).
It is a simple exercise to show that $\Omega^\prime$
 as given by eq.(2.7) satisfies
the equation
$${\cal D}\Omega^\prime={\cal D}\Omega-{\pd\hat V\over{\pd\p}}
\Bigm\vert_{\p=v}(\gamma v+{\cal D}v)\eqno(2.8)$$
so that indeed $\Omega^\prime   $ satisfies eq.(2.4) since by definition
$${\pd\hat V\over{\pd\p}}\Bigm\vert_{\p=v}={\pd V\over{\pd\p}}
\Bigm\vert_{\p=v}=0.\eqno(2.9)$$
Note that this choice of $\Omega^\prime$ corresponds
 to setting the cosmological constant
to zero order by order in perturbation theory.

A third possibility which is relevant to some recent work of Kastening
\REFS\KA{B.Kastening, \plb283B (1992) 287}
\REFSCON\KAS{B.Kastening, preprint UCLA/92/TEP/26}.\refsend
is to choose
$$\hbox{iii)}\qquad \Omega^\prime=\Omega^\prime(\l_i).\eqno (2.10)$$
That is, to choose $\Omega^\prime $ to be independent of $\mu$.
To leading order $\Omega^\prime$ is therefore obtained by solving the
equation
$$\b_i{\pd\Omega^\prime\over{\pd\l_i}}=-\m{\pd\Omega\over{\pd\m}}=
{1\over{32\pi^2}}STr\, M^4\Bigm\vert_{\p=0}\eqno(2.11)$$
where $STr$ is a spin-weighted trace and $M^2$ is the mass matrix for the
quantum fields as a function of $\phi$. In section (5) we will
construct the solution to eq.(2.11) for the $O(N)$ scalar case and compare the
result with Ref.~\KAS .

\line{\hfil}
\line{\bigbf 3. Solutions to the renormalisation group equation.\hfil}
\line{\hfil}

In this section we consider the solution to various forms of the
\RG equation for $V$, and show how these solutions can be used to extend the
domain of perturbative believability (in $\phi$) of the result: or
equivalently, sum the leading (and subleading...) logarithms.
We suppose that $V$ satisfies the equation
$${\cal D}V-\gamma\p{\pd V\over{\pd \p}}=0.\eqno(3.1)$$
Straightforward application of the method of characteristics leads to the
solution
$$V(\m,\l_i,\p)=V\bigl(\m(t),\l_i(t),\p(t)\bigr)\eqno(3.2)$$
where
$$\eqalignno{\m(t)&=\m e^t&(3.3)\cr
\p(t)&=\p\xi(t)&(3.4)\cr}$$
and
$$\xi(t)=\exp\biggl( -\int^t_0 \gamma\bigl(\l_i(t^\prime)\bigr)\, dt^\prime
\biggr).
\eqno(3.5)$$
$\lambda_i(t)$ are the usual running couplings and masses, determined
by the equations
$${d\l_i(t)\over{dt}}=\b_i\bigl(\l(t)\bigr)\eqno(3.6)$$
subject to the boundary conditions $\lambda_i(0)=\lambda_i$.
It is sometimes more convenient to use dimensional analysis to recast
eq.(3.2) as follows:-
$$ \bar{\cal D}V-4\bar\gamma V=0\eqno(3.7)$$
where
$$\bar{\cal D}=\m{\pd\over{\pd\m}}+\bar \b_i{\pd\over{\pd\l_i}}
\eqno(3.8)$$
and
$$\eqalign{\bar\b_i&=(\b_i+\delta_i\l_i\gamma)/(1+\gamma)\cr
\bar\gamma&=\gamma/(1+\gamma).}\eqno(3.9)$$
Here $\delta_i$ is the
 dimension of the coupling $\lambda_i$.
The solution of eq.(3.7) is
$$V(\m,\l_i,\p)=\bar\xi(t)^4V(\m(t),\bar\l_i(t),\p)\eqno(3.10)$$
where $\mu(t)$ is as in eq.(3.3). $\bar\xi(t)$ and $\bar\l_i(t)$ are
defined as in eq.(3.5) and (3.6) but with $\gamma\rightarrow\bar\gamma$,
$\b\rightarrow\bar\b$ and $\l(t)\rightarrow\bar\l(t)$. The absence of a
$\pd /\pd \p$ from eq.(3.7) accounts for the fact that $\p$ rather
than $\p(t)$ appears on the right-hand side of eq.(3.10).
Either form of the solution may be employed with equivalent
results; let us focus for the moment on eq.(3.10). Let us denote
$V(\mu(t),\bar\lambda_i(t),\phi)$ as $V(t,\phi)$ for short.
Now suppose we wish to calculate $V(\mu,\lambda_i,\phi)$ (
$\equiv V(0,\p)$)
for some $\mu,$ say, 100~GeV. The key to the usefulness of the \RG
is that we can choose a value of $t$ such that the perturbation series
for $V(t,\phi)$ converges more rapidly (for certain $\phi$) than the
series for $V(0,\phi)$. Moreover, there is nothing to stop us choosing a
different value of $t$ for each value of $\phi$. Now the perturbation
series for $V$ is characterised at large $\phi$ by powers of the
parameter $\lambda\ln(\phi/\mu)$ where $\lambda$ is some
dimensionless coupling. Then clearly perturbation theory is improved if
we choose $t$ such that $\m(t)\sim\p$,
as long as $\lambda(t)$ remains small. The precise domain of applicability
of the solution for a given choice of $t$ depends on the details of the
theory: in section (5)  we will consider in detail the case of $O(N)$
$\phi^4$ theory.

  Meanwhile, however, let us consider the relevance of the above
discussion to the issue of the subtraction term $\Omega^\prime(\m,\l_i)$
introduced in the previous section. The important point we wish to make
here is that whichever procedure we use to define $\Omega^\prime$,
and whether we use the \RG solution eq.(3.2) or eq.(3.10), a choice
of $t$ dependent on $\phi$ renders $\Omega^\prime$
a function of $\phi$ and hence no longer a trivial subtraction.
This point has been missed in some previous treatments of the \RG
solution and is implicit in the treatment of Kastening.

It is evident that, with regard to extending the domain of perturbative
calculability, one must take into account the behaviour of $\Omega^\prime
\bigl(\m(t),\l_i(t)\bigr)$
although, since it depends on $\phi$ only through $t$, this is unlikely to
pose a problem at large $\phi$, for example. But we can, in fact,
finesse this issue altogether by beginning with the \RG equation for
$V^\prime\equiv\pd V/\pd \p$
instead of the one for $V(\p)$,
the point being that
$${\pd V\over{\pd\p}}\bigl(\m,\l_i,\p\bigr)=
{\pd\hat V\over{\pd\p}}\bigl(\m,\l_i,\p\bigr)\eqno(3.11)$$
so that the $\Omega^\prime    $ term simply does not arise.
The analog to eq.(3.1) is
$${\cal D}V^\prime-\gamma\p{\pd V^\prime\over{\pd\p}}=\gamma V^\prime
\eqno(3.12)$$
with solution
$$V^\prime(\m,\l_i,\p)=\xi(t)V^\prime\bigl(\m(t),\l_i(t),\p(t)\bigr)
\eqno (3.13)$$
while the analog to eq.(3.10) is simply
$$V^\prime(\m,\l_i,\p)=\bar\xi(t)^4 V^\prime\bigl(
\m(t),\bar\l_i(t),\p\bigr)\eqno(3.14)$$
since $V^\prime $ evidently obeys an \RG equation of the same form as
eq.(3.7).

\line{\hfil}
\line{\bigbf 4. $\phi^4$ theory: the N=1 case.\hfil}
\line{\hfil}

In this section we apply the formalism developed in the previous two
sections to the case of massive $\lambda\phi^4$ theory, defined by the
Lagrangian
$${\cal L}={1\over2}(\pd_\m\p)^2-{1\over2}m^2\p^2-{\l\over{24}}\p^4.
\eqno(4.1)$$
$\hat V(\p)$ is given by the loopwise expansion
$$\hat V(\p)=\hat V_0+\hat V_1+\hat V_2+...\eqno (4.2)$$
where
$$\hat V_0={1\over2}m^2\p^2+{\l\over{24}}\p^4\eqno(4.3)$$
and
$$\hat V_1=\k{1\over4}H^2\Bigl(\ln {H\over{\m^2}}-{3\over2}
\Bigr).\eqno(4.4)$$
In eq.(4.4), $H=m^2+{1\over2}\lambda\phi^2$, $\k \equiv (16\pi^2)^{-1},$
and we are using $ \overline{MS}$
as we do throughout. (The result for $\hat V_2$ may be found in
\REF\FJ{C.Ford and D.R.T.Jones, \plb274B (1992) 409; erratum-
{\it ibid\/} 285B (1992) 399}~Ref.~\FJ.)

At the one loop level the relevant \RG functions are given by
$$\b_\l^{(1)}=3\l^2\k,\qquad\b_{m^2}^{(1)}
=m^2\l\k,\qquad\gamma^{(1)}=0\eqno(4.5a,b,c).$$
By virtue of eq.(4.5c) the two forms of the \RG solution are identical,
and we have
$$\eqalign{V(\m,\l,m^2,\p)=&\Omega^\prime\bigl(\m(t),\l(t),m^2(t)\bigr)
+{1\over2}m^2(t)\p^2+{1\over{24}}\l(t)\p^4\cr
&+{\k\over4}H^2(t)\Bigl(\ln{H(t)\over{\m^2(t)}}-{3\over2}\Bigr)
+...\cr}\eqno(4.6)$$
 where $H(t)=m^2(t)+{1\over 2}\lambda(t)\phi^2$,
$$\l(t)=\l(1-3\l t\k)^{-1}\eqno(4.7)$$
and
$$m^2(t)=m^2(1-3\l t\k)^{-1/3}.\eqno(4.8)$$
The function $\Omega^\prime$ depends
 on the choice made to achieve a $V$ satisfying the \RG
equation as explained in section (2). With choice (iii), ie
$\Omega^\prime(\m,\l,m^2)=\Omega^\prime(\l,m^2)$
   it is easy to show using eq.(2.11) that
$$\Omega^\prime(\l,m^2)=-{m^4\over{2\l}}+cm^4\l^{-2/3}.\eqno(4.9)$$
where $c$ is an arbitrary constant. Notice, that when $m^2$ and $\lambda$
become $t$-dependent in accordance with eq.(4.6)-(4.8) the $c$-term above
remains $t$-independent and therefore harmless; so we may set $c=0$.
Note that this choice of $\Omega^\prime$  has the curious feature that in the
free field limit ($\l\rightarrow 0$) it corresponds to an {\it infinite\/}
vacuum subtraction.
 We will return later to the consequences of choice (ii) for $\Omega^\prime$;
for the time being let us persist with eq.(4.9).
With this $\Omega^\prime$, in fact, eq.(4.6) essentially reproduces the
leading logarithms sum of Kastening (eq.(25) of Ref.~\KAS) . The
natural choice of $t$ from the point of view of eq.(4.6) is given by the
equation
$$\m^2(t)=\m^2e^{2t}=m^2(t)+{1\over2}\l(t)\p^2\eqno(4.10)$$
since this evidently removes the $\ln(H/\mu^2)$ terms to all orders.
An alternative choice which enables us to make contact with Kastening's work
is to choose\footnote\dag{For the purposes of this discussion we found
it convenient to write in the factors of ${\hbar}$ explicitly.}
$$
\m^2(t)=\m^2e^{2t/\hbar}=m^2+{1\over2}\l\p^2 \eqno(4.11)
$$
which is a less implicit definition of $t$ inasmuch as now
$$
t={\hbar\over2}\ln{m^2+{1\over2}\l\p^2\over{\m^2}}. \eqno(4.12)
$$
Now we show how the various leading logarithm (subleading logarithm....)
sums collected in Kastening's functions $f_1$, $f_2$ etc. are in fact
subsumed in our solution. (We choose now to work with eq.(3.2) rather
than eq.(3.10).) We need to expand the solution
$V(\m(t), \l(t), m^2(t), \p(t))$
in powers of $\hbar$
{\it but retaining all orders\/} in $t$. Thus, from the
expression for $\b_\l$ incorporating two-loop corrections:
$$
{d\l(t)\over{dt}} = 3\l^2(t)\k - {17\over3}{\hbar}\l^3(t)\k^2 +.....
\eqno(4.13)
$$
it is easy to show that
$$\l(t)=\l(1-3\l t\k)^{-1}+{17\over9}\hbar\l^2\k(1-3\l t\k)^{-2}\ln
(1-3\l t\k)+O(\hbar^2).\eqno(4.14)$$
Similarly we can evaluate $m^2(t)$, $\p(t)$ and $\xi(t)$ through two loops.
The relevant two-loop contributions to the \RG functions are
$$\b^{(2)}_\l=-{17\over3}\hbar^2\l^3\k^2,\qquad \b^{(2)}_{m^2}=-{5\over6}m^2
\hbar^2
\l^2\k^2,\qquad \gamma^{(2)}={1\over12}\hbar^2\l^2\k^2.\eqno(4.15a,b,c)$$
Using these results we get
$$\eqalign{m^2(t)=&m^2(1-3\l t\k)^{-1/3}\cr &+\hbar m^2(1-3\l t\k)^{-4/3}
\Bigl[{17\over{27}}\k\l\ln(1-3\l t\k)+{19\over{17}}\l^2t\k^2\Bigr]
+O(\hbar^2)\cr
\p(t)=&\p-{1\over{12}}\hbar \l^2 t\k\p(1-3\l t\k)^{-2}+O(\hbar^2)\cr
\xi(t)=&1-{1\over{12}}\hbar\l^2 t\k(1-3\l t\k)^{-2}+O(\hbar^2).\cr}
\eqno(4.16)$$
Using the formulae for $\l(t)$, $m^2(t)$, $\p(t)$ and $\xi(t)$ together
with eq.(3.2) or (3.13) one can sum the leading (subleading...)
logarithms in $V(\p)$ or $V^\prime(\p)$ respectively.
The sum of the leading logarithms is given by the $\hbar^0$ term in (3.2):
$$L_1={1\over2}m^2\p^2(1-3\l t\k)^{-1/3}+{1\over{24}}\l\p^4
(1-3\l t\k)^{-1}-{m^4\over{2\l}}(1-3\l t\k)^{1/3}.\eqno(4.17)$$
With $t$ defined as in eq.(4.12) this is identical to the result of
Ref.~\KA. To sum the subleading
logarithms one simply takes the $O(\hbar)$ contribution to (3.2).
(Note that we would need to calculate the one loop contribution to
$\Omega^\prime$).

We have gone through this exercise to demonstrate how the results of
Refs.~\KA,\KAS~may be recovered directly from
the solution of the {\it RG\/} equation. The analysis
is founded on choice (iii) for $\Omega^\prime$, which, as
we have already indicated, we find somewhat artificial, particularly with
regard
to the free field limit. In addition, in more complicated theories with many
couplings
the determination of the $\Omega^\prime(\l_i)$ satisfying eq.(2.11) becomes
onerous.
We could choose to adopt choice (ii); it is easy to see, however, that the
result will then include terms of the form ${H^\prime}^2\ln{H^\prime/\m^2(t)}$
where $H^\prime = m^2(t)+{1\over 2}\l(t)\langle\p\rangle^2$. Although such
terms are not
dangerous at large $\p$ since they do not grow as  $\p^4$, they do lead to
an unwieldy form of the solution. With a view to more complicated theories
, it appears to us simpler, as we
indicated already, to work with $V^\prime=\pd V/\pd\p$. Then through one
loop we have (from either eq.(3.13) or (3.14))  simply
$$V^\prime=m^2(t)\p+{1\over6}\l(t)\p^3+{\k\over2}\l(t)\p H(t)
\Bigl(\ln {H(t)\over{\m^2(t)}}-1\Bigr)+...\eqno(4.18)$$
\indent We now evaluate $V^\prime$ and hence (numerically) $V$ with $t$
defined as in eq.(4.10).(Note that since $t$ depends
nontrivially on $\p$, the result for $V$ differs from that obtained
from the equivalent \RG equation for $V$ itself).
For $m^2>0$ and sufficiently small $\l$, the
result differs insignificantly from the tree result for
$\p<O(\m e^{1/\l\k})$, which corresponds to the approach of $\l(t)$ to
the Landau pole. For $m^2<0$ there is the fact that for
$H=m^2+{1\over 2}\l\p^2<0$ the ``unimproved" potential develops an
imaginary part, and there is no solution for $t$ to eq.(4.10).
Discussion of the imaginary part notwithstanding, it is clear that
perturbation theory is not to be trusted for $H\rightarrow 0$, as
follows. If we consider the higher order graphs constructed from the cubic
interaction only, then using dimensional analysis these contribute
to $V(\p)$ terms of the general form $(\l\p)^4\eta^{L-3}$ where
$$\eta={\k\l^2\p^2\over{m^2+{1\over2}\l\p^2}}\eqno(4.19)$$
and L is the number of loops.
Since $\eta\rightarrow\infty$ as $H\rightarrow 0$ we clearly have
perturbative breakdown in this region. This sort of infra-red
problem is characteristic of super-renormalisable interactions and is
important, of course, in calculations of $V$ at finite temperature.
Note that in the neighbourhood of the tree minimum,
$m^2+{1\over6}\l\p^2\approx 0$, we have $\eta\sim \l$ so
perturbative calculability requires merely $\k\l(t)\ll 1$ as we have
already assumed.

Finally let us consider briefly the massless case, $m^2=0$. As originally
indicated by {\it CW\/}, $V$ then remains well defined and perturbatively
calculable for $\p\rightarrow 0$, so that $\p=0$ remains a local
minimum (and the global one, modulo the fact that as before $V$ can not
be calculated in the neighbourhood of the Landau pole).

\line{\hfil}
\line{\bigbf 5. O(N) $\p^4$ theory.\hfil}
\line{\hfil}

Here we generalise section (4) to the case of massive $O(N)$ symmetric
$\p^4$ theory, defined by the Lagrangian
$${\cal L}={1\over2}(\pd_\m\vec\p)^2-{1\over2}m^2{\vec\p}^2
-{\l\over{24}}({\vec\p}^2)^2\eqno(5.1)$$
where ${\vec\p}^2=\sum^N_{i=1}\p^i\p^i$.
Including one loop corrections the effective potential is given by
$$\eqalign{V(\p)=&\Omega^\prime+{1\over2}m^2\p^2+{\l\over{24}}\p^4\cr
&+{\k\over4}H^2\Bigl(\ln {H\over{\m^2}}-{3\over2}\Bigr)
+{\k\over4}(N-1)G^2\Bigl(\ln {G\over{\m^2}}-{3\over2}\Bigr)\cr}
\eqno(5.2)$$
where $G=m^2+\l\p^2/6$, and we have exploited the $O(N)$ invariance
to write $V$ as a function of a single field $\p$. Once again the
two-loop corrections may be found in Ref.~\FJ.

At the one loop level the relevant  \RG functions are
$$\b_\l^{(1)}={N+8\over3}\l^2\k,\qquad \b_{m^2}^{(1)}
={N+2\over3}m^2\l\k,\qquad
\gamma^{(1)}=0.\eqno(5.3a,b,c)$$
\indent As explained in previous sections, we prefer to deal with the \RG
equation
for $V^\prime$ but we note for completeness that if we choose to define
$V$ \`a la Kastening then writing $\Omega^\prime=m^4f(\l)$ we have from
eq. (2.11) that
$$\l{df\over{d\l}}+2{N+2\over{N+8}}f={3N\over{2(N+8)\l}}\eqno(5.4)$$
with solution
$$
f=\cases{3N[2(N-4)\l]^{-1}+c\l^{-2(N+2)/(N+8)},&if $N\ne4$\cr
(\ln\l)/(2\l)+c/\l,&if $N=4$\cr}\eqno(5.5)
$$
As in the previous section the $c$-terms in $\Omega^\prime$ are in
fact $t$-independent in the  \RG solution so we may set $c=0$.
It is easy to see that for $N\ne4$ eq.(5.5) corresponds to eq.(15)
of Ref.~\KAS~(with $t=0$).

Reverting now to $V^\prime$, we have from eq.(3.13) that
$$\eqalign{V^\prime(\m,m^2,\l,\p)=&m^2(t)\p+{1\over6}\l(t)\p^3\cr
&+{\k\over2}\l\p H(t)\Bigl(\ln {H(t)\over{\m^2(t)}}-1\Bigr)
+{\k\over6}(N-1)\l\p G(t)\Bigl(\ln {G(t)\over{\m^2(t)}}-1\Bigr)+....
\cr}\eqno(5.6)$$
Evidently there is no choice of $t$ which eliminates the logarithms to
all orders: but if our concern is to control the behaviour of $V$
at large $\p$ then any choice such that $\m^2(t)\sim \p^2$ will do.
With (say) $t=\ln(\p/\m)$, it is a simple matter to compute
$V^\prime$ as defined by eq.(5.6) and hence (numerically)
$V(\m,m^2,\l,\p)$. For $\k\l\ll 1$, the result differs little from the tree
approximation out to $\p\sim \m e^{1/(\k\l)}$ just as in the $N=1$
case.

As in the $N=1$ case perturbation theory will break down (for $m^2<0$)
in the region $H\approx 0$. We now, however, have also to consider
whether there are also IR problems at $G\approx 0$: ie at the tree
minimum. Evidently for $G<0$, $V$ becomes complex: but how closely can
we approach $G=0$ from above and retain perturbative calculability?
In fact there is no problem as $G\rightarrow 0$; this is evident
explicitly at one and two\refmark\FJ\ loops. To extend this result to higher
loops,
note that we have in general cubic vertices of the type $H^3$ and
$HGG$ but not $G^3$. Consider some graph consisting of $HGG$ vertices
only: if it is singular as $G\rightarrow 0$, then it will still be
so if we ``shrink" every $H$ propagator by the substitution
$1/(k^2+H)\rightarrow 1/H$. But the diagram will then consist of $G^4$
vertices only, with the effective coupling $\l^2\p^2/H$. Then by
dimensional analysis, or simply by noting that $G^4$ is a
renormalisable (not a super-renormalisable) vertex, it is clear that the
graph will not be singular as $G\rightarrow 0$. The significance of
the fact that
$\pd^2 V/\pd \p^2$ is singular at $G=0$ is not precisely clear to us;
at the true minimum, of course,
(calculated consistently to any order in $\hbar$) the matrix
$\pd^2 V/\pd \p^i \pd \p^j$ has no singularities and N-1 zeroes
corresponding to the would-be Goldstones.

\line{\hfil}
\line{\bigbf 6. The standard model.\hfil}
\line{\hfil}

In this section we consider $V(\p)$ in the standard model (\SM) from the \RG
point of view, with emphasis on the question of vacuum stability. As in
the $O(N)$ scalar case we can exploit gauge invariance to write $V$ as
a function of a single field $\p$.
We must also choose a gauge; the 't Hooft-Landau gauge is the most
convenient. In this gauge the $W$, $Z$ and $\gamma$ are transverse, and the
associated ghosts are massless and couple only to the gauge fields;
 the would be Goldstone bosons $G^{\pm}$,$G$ have a common mass
deriving from the scalar potential only. Moreover, the gauge parameter
is not renormalised in this gauge so it does not enter
the \RG equation.

Calculating $V$ through one loop yields
$$
\eqalign{V(\p)=&\Omega^\prime(\m,m^2,h,\l,g,g^\prime)+
{1\over2}m^2\p^2+{1\over{24}}\l\p^4\cr
&+{\k}\Bigl[ {1\over4}H^2\Bigl(\ln {H\over{\m^2}}-
{3\over2}\Bigr)+{3\over4}G^2\Bigl(\ln{G\over{\m^2}}-{3\over2}\Bigr)
+{3\over2}W^2\Bigl(\ln{W\over{\m^2}}-{5\over6}\Bigr)\cr
&+{3\over4}Z^2\Bigl(\ln{Z\over{\m^2}}-{5\over6}\Bigr)
-3T^2\Bigl(\ln{T\over{\m^2}}-{3\over2}\Bigr)\Bigr]+...\cr}\eqno(6.1)$$
where
 $$H=m^2+{1\over2}\l\p^2,\qquad T={1\over2}h^2\p^2,
\qquad G=m^2+{1\over6}\l\p^2,$$
$$W={1\over4}g^2\p^2,\qquad Z={1\over4}(g^2+{g^\prime}^2)\p^2.
$$
Here $h$ is the top quark Yukawa coupling (we neglect other
Yukawa couplings throughout).

The occurrence of the logarithms of $H$,$G$,$T$,$W$ and $Z$ in the
perturbation expansion means of course that no choice of $t$ will
eliminate the logarithms altogether. As indicated in the $O(N)$ scalar
case, however, it is clear that as long as the initial values of the
dimensionless couplings are small and they remain small on evolution then  as
long as we choose $\m(t)\sim \p$, our \RG solution eq.(3.14), say, will be
perturbatively believable for all $\p$.

The essential feature that distinguishes gauge theories in general from the
pure scalar cases discussed in the previous two sections is the fact that
$\l=0$ is no longer a fixed point in the evolution of the quartic scalar
coupling $\l(t)$. Evolution of $\l$ with $\p$ may therefore drive $\l$
negative and hence cause $V$ to develop a second local minimum\footnote\dag
{If one chooses to identify this ``new" minimum with the true electroweak
vacuum then it is easy to see that this results in the ``Coleman-Weinberg"
vacuum with a concomitant experimentally disfavoured prediction for
the Higgs mass.\refmark\Sher}
 at large
$\p$; if this minimum is deeper than the (radiatively corrected) tree
minimum then it will result in the destabilisation of the electroweak
vacuum. Requiring stability (or at least longevity) of the electroweak
vacuum results in an upper limit on $m_t$ (for a given $m_H$). The
existence of this limit and related issues has been explored in a series
of papers by Sher et al\refmark\Shers\ (for a clear and comprehensive review
see
Ref.~\Sher ).

Now (as in fact essentially recognised by Sher in Ref.~\Sher)~the form of the
\RG ``improved" $V$ used in Ref.~\Shers~is not completely satisfactory,
inasmuch as it is not in general a solution of the \RG equation
for all values of the Higgs $(mass)^2$ parameter $m^2$.
In fact, however, because the false minimum, if present, occurs at large $t$
(and hence $\p\gg M_Z$) this should make little difference. Provided a choice
of $t$ is made such that $\m(t)\sim\p$ (at large $\p$),
contributions to $V$ from the $\Omega^\prime$ term and subleading logarithms
neglected in Ref.~\Sher~are very small for values
of $m_t$ and $m_H$ in the range
of interest. In fact it is easy to convince oneself that in terms of the
 solution eq.(3.10), for example, the question of the existence of a false
(deep) minimum at some scale is simply the question of whether $\l(t)$
goes negative as $t$ increases. Even for very small negative $\l$, the fact
that this happens at $\p/M_Z\gg 1$ means that the tree term $\l\p^4/24$ drives
$V$ well below the electroweak minimum. Thus although we now have
available the two-loop corrections to $V$\refmark\fij\  for the \SM, they will
have a
negligible affect on the outcome. The importance of the evolution of
$\l$ to the stability of the vacuum was in fact recognised in~
\REF\cab{N.Cabibbo, L.Maiani, G. Parisi and R.Petronzio, \npb158 (1979) 295}
Ref.~\cab~and the calculation performed using the one loop \SM beta functions.
The main question we resolve in this section is  the effect of
2 loop corrections on this calculation. (Previous calculations of this
 correction are unreliable due to typographical error in the expression for
$\b_\l^{(2)}$ given in
\REF\MV{M.E.Machacek and M.T.Vaughn, \npb222 (1983) 83}~Ref~\MV.)~
In fact we have also calculated the
evolution of $m^2$ through two loops and hence the improved $V$
 as a function of $\p$ but, as anticipated above, the
requirement that the electroweak vacuum remains stable turns out to
essentially identical to the requirement that $\l$ remains positive.

We give the \SM $\b$-functions through two loops in an appendix. It only
remains to discuss boundary conditions. At $\m=M_Z$ we use input values
for $g$, $g^\prime$, $\alpha_3$, $\l$, $h$, $m^2$,
as follows:
$$\eqalign{g&=0.650\cr
g^\prime&=0.358\cr
\alpha_3&=0.10, 0.11, 0.12, 0.13\cr
\l&=\l_0\cr
h&=h_0\cr
m^2&=m^2_0.\cr}\eqno(6.2)$$
In order to translate the results into a limit on $m_t$, $m_H$ we use the
tree results
$$\eqalign{m_t&={1\over{\sqrt 2}}h_0v_0\cr
m_H^2&=-2m^2_0={1\over3}\l_0v^2\cr}\eqno(6.3)$$
where $v^2=-6m^2_0/\l_0$ ($=(246~{GeV})^2$). (Of course these relationships are
themselves subject to radiative corrections which we could include in
principle).

Because the $-36h^4$ term in $\b_\l^{(1)}$ tends to drive $\l$ negative, the
result of the evolution is a lower limit on $\l_0$ (and hence $m_H$) for
a given $h_0$ (and hence $m_t$). Now the evolution equation for $h$ ( see
eq.(A1)) includes a contribution from $\alpha_3$; increasing the
input value of $\alpha_3$ causes $h$ to decrease faster as $t$ increases,
and so we would expect the lower bound on $m_H$ to {\it decrease\/} with
increasing $\as$.

In fig(1) we display the evolution of $\l$ against $t$ for $m_t=120~GeV$
and three values of $\l_0$. For $\l_0\approx 0.120$, $\l(t)$ goes negative but
remains small and becomes
positive again for $t\sim 15$; but nevertheless because it is negative
(albeit small) for $t\sim 10$ this results in a very deep minimum at
large $\p$. The value $\l_0=0.125$ is the critical value, corresponding
to $m_H=50.3~GeV$.

In fig.(2) we display the critical $m_H$ as a function of $m_t$ for
$\alpha_3=0.11$, as obtained in the one- and two-loop approximations,
respectively. We see that the two-loop corrections are not very large;
typically they decrease the lower bound on $m_H$ by $2-4~GeV$ or so.

In fig.(3) we present the critical curve for four input values of
$\as$ . The dependence on $\as$ is quite marked, and as anticipated above,
the lower bound on $m_H$ decreases as $\as$ increases. This conclusion
is at variance to that of Ref.~\Sherz, where the sensitivity to $\as$
was indeed noted, but the bound on $m_H$ was found to {\it increase\/}
as $\as$ increases.\footnote{*}{We thank Marc Sher for confirming that the
lower bound on $m_H$ indeed decreases as $\alpha_3 (M_Z)$
increases; the result of Ref.~\Sherz~ was due to a printing error.}
We find, for example that for $m_t=130~GeV$,
the bound on $m_H$ is given by $70.1~GeV$ if $\as = 0.1$,
but $59.6~GeV$ if $\as = 0.13$.

For $m_t\geq 140~GeV$ the curves are to a very good approximation linear,
and stability of the electroweak vacuum corresponds in this region
to the relationship ( for $\as =0.11$, for example)
$$m_H\geq 1.95m_t-189~GeV.\eqno(6.4)$$
This differs somewhat from the linear approximation
 given by Sher (Ref.~\Sher~p331), which corresponds to
$m_H\geq 1.7m_t-160~GeV$. The reason for this discrepancy is that the latter
result is based on an extrapolation of the results for lower Higgs masses.
\footnote{**}{Once again we thank Marc Sher for confirming this.}

Let us consider briefly our results in the light of recent predictions
\Ref\efl{J.Ellis, G.L.Fogli and E.Lisi, preprint CERN-TH.6568/92}~for
$m_t$ and $m_H$ based on analysis of LEP data including radiative corrections:
$$
m_t=124^{+26}_{-28}~GeV \eqno(6.5)
$$
and
$$
m_H=25^{+275}_{-19}~GeV.\eqno(6.6)
$$
With $m_t=120~GeV$, for instance, we have from fig.(3) that (again with
$\as = 0.11$)
$
m_H\geq 50.3~GeV
$
( $52.6~GeV$ from a one loop analysis).
So with this value of $m_t$ we are already
assured of vacuum stability by the {\it direct search\/} limit on $m_H$,
$m_H\geq~59 GeV$. For $m_t=140~GeV$, we have from fig.(3)
that $m_H\geq 83.2~GeV$.
Discovery of the Higgs (with this value of $m_t$) in the interval
$59~GeV\leq m_H \leq 83~GeV$ would strongly suggest the existence of
physics beyond
the standard model, since the obvious means to rescue electroweak stability
would be by new
physics at a scale heavy enough to have negligible impact on the radiative
corrections responsible for the results eq.(6.5) and (6.6). It is also clear
that refinement of the value of $\as (M_Z)$ would be helpful in reducing
the uncertainty in the critical curve.

\line{\hfil}
\eject
\line{\bigbf 7. Conclusions.\hfil}
\line{\hfil}

The renormalisation group expresses the simple fact that observables are
independent
of the renormalisation scale $\m$. Consequently, adroit choice of $\m$ leads
to improved perturbation theory by removing large logarithms in processes
characterised by a single momentum scale.\footnote{\dag}{Processes with
several scales may benefit from a multiscale {\it RG\/} approach: see
\REF\EJMS{M.B.Einhorn and D.R.T.Jones, \npb230 (1984) 261}~Ref~\EJMS.}
Application of the {\it RG\/} to the effective potential is quite analagous,
except now it is the region of large (or small) $\p$ that becomes accessible.
In this paper we hope we have elucidated the issues that arise; in
particular the relationship between the usual \RG approach and the analysis
of Ref.~{\KA , \KAS}. We have also reconsidered the \RG improvement of the \SM
 potential, and give a result for the electroweak stability bound on $m_H$
based on a full two loop \RG analysis. In particular we highlighted
the dependence on $\as (M_Z)$, showing that the lower bound on $m_H$ decreases
with increasing $\as(M_Z)$. With the discovery of the
top quark generally expected to be imminent, it will be interesting
to see whether the direct search limit on $m_H$ leaves a ``window of
instability",
as discussed in section 6.

Among further applications of the \RG to the effective potential, we
might consider extension to the supersymmetric \SM, and also whether the \RG
improved potential has any bearing on the issue of triviality of
non-asymptotically
free theories.

\line{\hfil}
\line{\bigbf Acknowledgements.\hfil}
\line{\hfil}
 While part of this work was done, one of us (DRTJ) enjoyed the hospitality of
the Institute for Theoretical Physics at Santa Barbara and thanks Jim Langer
for his part in making the visit possible. C.F is grateful to the S.E.R.C for
financial support; this research was also supported by the National Science
Foundation under grant no. PHY89-04035, and by a NATO collaboration research
grant.  The work of one of us (MBE) was supported in part by the U.S.
Department of Energy and by the Institute for Theoretical Physics at Santa
Barbara.

\line{\hfil}
\eject
\line{\bigbf \hfil Appendix\hfil}
\line{\hfil}

We list the \RG functions for the \SM (see section 6 for notation
and conventions) through  two loops.

The one-loop \RG functions are
$$\eqalign{\ik\gamma^{(1)}=& 3h^2-{9\over4}g^2-{3\over4}{g^\prime}^2\cr
\ik\b_\l^{(1)}=&4\l^2+12\l h^2-36h^4-9\l g^2-3\l {g^\prime}^2\cr
&+{9\over4}{g^\prime}^4+{9\over2}g^2{g^\prime}^2+{27\over4}g^4\cr
\ik\b_h^{(1)}=&{9\over2}h^3-8g_3^2 h-{9\over4}g^2 h-{17\over{12}}
{g^\prime}^2h\cr
\ik\b_g^{(1)}=&-{19\over6}g^3\cr
\ik\b_{g^\prime}^{(1)}=&{41\over6}{g^\prime}^3\cr
\ik\b_{g_3}^{(1)}=&-7g_3^3\cr
\ik\b_{m^2}^{(1)}=&m^2(2\l+6h^2-{9\over2}g^2-{3\over2}{g^\prime}^2).\cr}\eqno(A1)
$$
The two-loop contributions to the \RG functions are given by
$$\eqalign{\il\gamma^{(2)}=&{1\over6}\l^2-{27\over4}h^4+20g_3^2h^2
+{45\over8}g^2h^2+{85\over{24}}{g^\prime}^2h^2\cr
&-{271\over{32}}g^4+{9\over{16}}g^2{g^\prime}^2+{431\over{96}}
{g^\prime}^4\cr
\il\b_\l^{(2)}=&-{26\over3}\l^3-24\l^2h^2+6\l^2(3g^2+{g^\prime}^2)
-3\l h^4+80\l g_3^2h^2\cr
&+{45\over2}\l g^2h^2+{85\over6}\l{g^\prime}^2h^2-{73\over8}\l g^4
+{39\over4}\l g^2{g^\prime}^2+{629\over{24}}\l{g^\prime}^4\cr
&+180h^6-192h^4g_3^2-16h^4{g^\prime}^2-{27\over2}h^2g^4
+63h^2g^2{g^\prime}^2\cr
&-{57\over2}h^2{g^\prime}^4+{915\over8}g^6-{289\over8}g^4{g^\prime}^2
-{559\over8}g^2{g^\prime}^4-{379\over{8}}{g^\prime}^6\cr
\il\b_h^{(2)}=&h\bigl(-12h^4+h^2({131\over{16}}{g^\prime}^2
+{225\over{16}}g^2+36g_3^2-2\l)+{1187\over{216}}{g^\prime}^4\cr
&-{3\over4}g^2{g^\prime}^2+{19\over9}{g^\prime}^2g_3^2-{23\over4}g^4
+9g^2g_3^2-108g_3^4+{1\over6}\l^2\bigr)\cr
\il\b_g^{(2)}=&g^3({3\over2}{g^\prime}^2+{35\over6}g^2+12g_3^2-{3\over2}h^2)\cr
\il\b_{g^\prime}^{(2)}=&{g^\prime}^3({199\over{18}}{g^\prime}^2
+{9\over2}g^2+{44\over3}g_3^2-{17\over6}h^2)\cr
\il\b_{g_3}^{(2)}=&g^3_3({11\over6}{g^\prime}^2+{9\over2}g^2-26g_3^2-2h^2)\cr
\il\b_{m^2}^{(2)}=&2m^2\bigl(-{5\over6}\l^2-6\l h^2+2\l(3g^2+{g^\prime}^2)
-{27\over4}h^4+20g_3^2h^2\cr
&+{45\over8}g^2h^2+{85\over24}{g^\prime}^2h^2-{145\over{32}}
g^4+{15\over{16}}g^2{g^\prime}^2+{157\over{96}}{g^\prime}^4\bigl).\cr}\eqno(A2)
$$
\vfill\eject
\refout
\vfill\eject
\line{\bigbf FIGURE CAPTIONS\hfil}
\bigskip
\line{$\underline{Fig.1}$\hfil}
Plot of the running coupling $\l(t)$ for $m_t=120~GeV$
and $\l_0$ just above, at and just below its critical value $(0.125)$.
\bigskip
\line{$\underline{Fig.2}$\hfil}
Plot of the critical value of $m_H$ for vacuum stability
against $m_t$, for $\as (M_Z) =0.11$, showing one- and two-loop approximations.
\bigskip
\line{$\underline{Fig.3}$\hfil}
Plot of the critical value of $m_H$ for vacuum stability
against $m_t$, for $\as (M_Z) =0.1, 0.11, 0.12$ and $0.13$.
\end